# Phonon Theory of Martensitic Transformation Precursors

Yongmei M. Jin† and Yu U. Wang‡

Department of Materials Science and Engineering,
Michigan Technological University, Houghton, MI 49931, USA

**Abstract**

A phonon theory of precursor phenomena in martensitic phase transformations is developed. Extending Grüneisen theory of thermal expansion, this theory addresses the effects of deformation-dependent low-energy phonons on the structural, thermal, and elastic behaviors of pre-martensitic cubic crystals that undergo incomplete phonon softening. It reveals spontaneous symmetry breaking, pre-martensitic transformation, phonon domains, and tweed structure. The theory naturally explains the ubiquitous pre-transitional anomalies that are difficult to explain from conventional phase transition theories.

Martensitic transformation is a typical solid-state displacive phase transition that breaks the symmetry of crystal structure by development of spontaneous anisotropic lattice strain upon cooling [1,2]. The pre-martensitic high-symmetry phases usually undergo phonon softening in a wide temperature range above the transition temperatures [2,3]. While the high-symmetry phases are energetically unstable, the crystal structures are stabilized by the excess phonon entropy due to the lower energy of softened phonons [4,5]. Soft-mode theory [6], originally developed for ferroelectric phase transition, has been used to describe second-order martensitic transformations in some systems that undergo complete phonon softening. However, the great majority of martensitic transformations are first-order phase transitions, where certain phonon modes only exhibit incomplete softening and the soft-mode theory is inapplicable [7]. Phenomenological two-strain Landau theory has been developed for martensitic transformations, where no soft mode is needed and the coupling between uniform lattice strain and internal shuffling strain leads to first-order phase transition [8]. Anomalous precursor phenomena generally occur in cubic phases prior to the martensitic transformations, including diffuse scattering in diffraction, tweed patterns in transmission electron microscopy, and thermal, acoustic and elastic anomalies, which cannot be explained by the current phase transition theories [9,10]. A Landau theory introduces, in addition to the two strain order parameters, magnetization as a third order parameter and shows that magnetoelastic coupling drives a first-order pre-martensitic transition in magnetic shape memory alloys [11]. Since pre-transitional phenomena are generally observed in both non-magnetic and magnetic materials, a general theory is still needed to clarify the fundamental origin of the ubiquitous martensitic precursor phenomena.

In this Letter, we develop a phonon theory that provides a simple and natural explanation to the ubiquitous precursor phenomena in martensitic phase transformations. Extending Grüneisen theory of thermal expansion [12], this theory addresses the effects of deformation-dependent incompletely-softened low-energy phonons on the structural, thermal, and elastic behaviors of pre-martensitic cubic crystals. Previously unexplored fundamental phenomena distinct from the symmetry-preserving thermal expansion are revealed in the cubic phases, including spontaneous symmetry breaking (pre-martensitic transformation), phonon domains, tweed structures, diffuse scattering, and thermal, acoustic and elastic anomalies. It shows that martensitic precursor phenomena are intrinsic behaviors (as natural as thermal expansion) of anharmonic lattice dynamics in cubic crystals that undergo incomplete phonon softening.

Due to the drastically distinct time scales of lattice deformation and thermal vibrations in a crystal, the internal coordinates of the solid can be separated into static (configurational) ones describing lattice strain and dynamic (vibrational) ones describing lattice vibrations about deformed equilibrium lattice sites. The free energy density $F$ of the crystal can be expressed as a sum of elastic free energy (with the constant free energy of the undeformed lattice omitted) and phonon free energy over all normal vibration modes labeled by $\{\mathbf{k},p\}$ (under independent harmonic oscillator approximation) [12]:

$$F = \frac{1}{2}C_{ijkl}\varepsilon_{ij}\varepsilon_{kl} + \frac{1}{V}\sum_{\mathbf{k},p}\left\{\frac{\hbar\omega_{\mathbf{k},p}}{2} + k_{\mathrm{B}}T\ln\left[1-\exp\left(-\frac{\hbar\omega_{\mathbf{k},p}}{k_{\mathrm{B}}T}\right)\right]\right\} \qquad (1)$$

where $C_{ijkl}$ is elastic stiffness tensor, $\varepsilon_{ij}$ is strain tensor, $\hbar$ and $k_{\mathrm{B}}$ are reduced Planck constant and Boltzmann constant, respectively, $\omega_{\mathbf{k},p}$ is angular frequency of phonon mode with wavevector $\mathbf{k}$ and polarization $p$, $V$ is crystal volume, and $T$ is thermodynamic temperature. It is worth noting that, while the free energy in Eq. (1) is formulated based on harmonic approximation, introduction of strain dependence to phonon frequencies makes the theory anharmonic, as shown in the following.

To illustrate spontaneous symmetry breaking of crystal lattice due to deformation-dependent incompletely-softened low-energy phonons, we consider loss of lattice stability in cubic crystal with respect to tetragonal distortion as an example. For convenience, we introduce volumetric strain $\varepsilon_V = \varepsilon_1 + \varepsilon_2 + \varepsilon_3$ and tetragonal strain $\eta = \varepsilon_1 - \varepsilon_3 = \varepsilon_2 - \varepsilon_3$. It is worth noting that $\varepsilon_V$ describes pure volume change without change in lattice symmetry, while $\eta$ characterizes tetragonal lattice distortion thus breaking of cubic symmetry, i.e., $\eta$ serving as an order parameter of displacive phase transformation. In Voigt notation of cubic elastic constants, the free energy in Eq. (1) becomes

$$F = \frac{1}{6}(C_{11} + 2C_{12})\varepsilon_V^2 + \frac{1}{3}(C_{11} - C_{12})\eta^2 + \frac{k_B T}{V} \sum_{\mathbf{k}}' \sum_{\{\mathbf{k}\}} \sum_p f(x_{\mathbf{k},p}) \quad (2)$$

where $\sum_{\mathbf{k}}'$ indicates sum over the wavevectors $\mathbf{k}$ only in irreducible first Brillouin zone (IFBZ), $\sum_{\{\mathbf{k}\}}$ indicates sum over all wavevectors $\{\mathbf{k}\}$ that are symmetry-related to the wavevector $\mathbf{k}$ in IFBZ (i.e., belonging to the same star $\{\mathbf{k}\}$ that is a set of wavevectors obtained from one wavevector $\mathbf{k}$ in IFBZ via all operations of the symmetry group of the cubic lattice [1]), $x_{\mathbf{k},p} = \hbar\omega_{\mathbf{k},p}/k_B T$ is dimensionless phonon energy, and

$$f(x) = \frac{x}{2} + \ln(1 - e^{-x}) \quad (3)$$

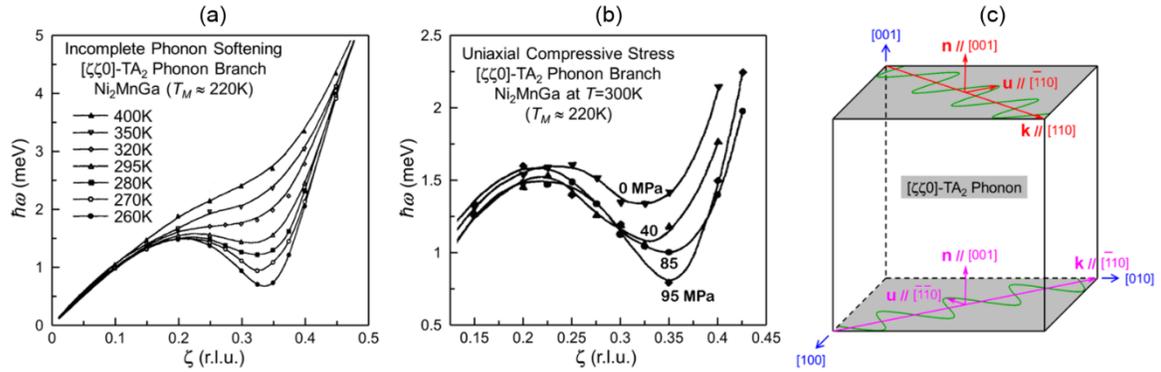

FIG. 1 (color online). Dependence of [ζζ0] transverse acoustic TA$_2$ phonon branch on (a) temperature and (b) uniaxial compressive stress in pre-martensitic Ni$_2$MnGa, as determined by inelastic neutron scattering experiments (reproduced with permission – Ref. 13 and 14, respectively). (c) Schematic illustration of wavevector $\mathbf{k}$, polarization $\mathbf{u}$ and normal $\mathbf{n}$ of [ζζ0]-TA$_2$ phonon modes in cubic crystal.

For solids whose phonon dispersions depend on deformation, $\omega_{\mathbf{k},p}$ is function of $\varepsilon_{ij}$, i.e., $\omega_{\mathbf{k},p}(\varepsilon_{ij})$. To investigate the effects of deformation-dependent phonons on the lattice stability of cubic crystals, only the phonon modes that both possess low energy and sensitively depend on strain are of importance in the summation in Eq. (2), as shown in the following. To be specific, we consider [ζζ0] transverse acoustic TA2 phonons which are relevant to most martensitic systems [3]. In cubic austenite phases 10-100 K above martensite start temperatures, this phonon branch undergoes incomplete softening, where the phonon dispersion curve develops a dip at $\zeta \approx 0.33$ and the corresponding low-energy phonons with $\hbar\omega \sim 1\,\text{meV}$ exhibit strong dependence on deformation, as exemplified in Fig. 1(a) [13] and (b) [14]. It is these deformation-dependent low-energy phonon modes that play a dominant role in the lattice stability of pre-martensitic cubic crystals, i.e., $\sum'_{\mathbf{k}} \sum_{p}$ sum only over a small part of IFBZ and only over [ζζ0]-TA2 modes. For simplicity without loss in conceptual generality, we adopt Einstein-type model to focus on $N'$ [ζζ0]-TA2 phonon modes around $\zeta \approx 0.33$ in first Brillouin zone (FBZ) ($N' \ll 3N$, $N$ being the total number of wavevectors in FBZ) sharing common frequency $\omega_0$ and strain dependence. This simplification allows an illustration of the basic features determined only by symmetry and a few material parameters. Figure 1(c) schematically illustrates [ζζ0]-TA2 phonon mode of wavevector $\mathbf{k}$ along [110], polarization $\mathbf{u}$ along [$\bar{1}$10] and perpendicular direction (hereafter called normal direction which determines the imminent tetragonal $c$-axis) $\mathbf{n}$ along [001] in cubic crystal. An equivalent phonon mode of the same $\mathbf{n}$ along [001] while $\mathbf{k}$ along [$\bar{1}$10] and $\mathbf{u}$ along [$\bar{1}\bar{1}$0] is also shown in Fig. 1(c). In this case, the star $\{\mathbf{k}\}$ consists of 12 equivalent [ζζ0]-TA2 phonon modes in cubic crystal, which fall into 3 groups according to $\mathbf{n}$ vector: $\mathbf{n}_1$//[100], $\mathbf{n}_2$//[010], $\mathbf{n}_3$//[001], each group consisting of four [ζζ0]-TA2 phonon modes characterized by $\mathbf{k}$ along four $\langle 110 \rangle$ directions within one $\{001\}$ plane. The phonon frequency $\omega_i$ of each group characterized by $\mathbf{n}_i$ exhibits the same dependence on normal strains. In terms of $x_i = \hbar\omega_i / k_B T$, the strain dependence of the 3 groups of [ζζ0]-TA2 phonon modes can be approximated in Taylor expansion with the symmetry of cubic crystal taken into account:

$$\begin{bmatrix} x_1 \\ x_2 \\ x_3 \end{bmatrix} = \begin{bmatrix} x_0 \\ x_0 \\ x_0 \end{bmatrix} + \begin{bmatrix} c & a & a \\ a & c & a \\ a & a & c \end{bmatrix} \begin{bmatrix} \varepsilon_1 \\ \varepsilon_2 \\ \varepsilon_3 \end{bmatrix} + \begin{bmatrix} g(\varepsilon_1, \varepsilon_2, \varepsilon_3) \\ g(\varepsilon_2, \varepsilon_3, \varepsilon_1) \\ g(\varepsilon_3, \varepsilon_1, \varepsilon_2) \end{bmatrix} \qquad (4)$$

where $x_0 = x_0(T)$ describes the temperature dependence of phonon energy $\hbar\omega_0(T)$, $a$ and $c$ are material parameters, and $g$ represents higher-order terms [15]. For tetragonal deformation characterized by $(\varepsilon_V, \eta)$, Eq. (4) yields

$$x_1 = x_2 = x_0 + \frac{1}{3}(c+2a)\varepsilon_V + \frac{1}{3}(c-a)\eta + h_1(\varepsilon_V, \eta) \qquad (5a)$$

$$x_3 = x_0 + \frac{1}{3}(c+2a)\varepsilon_V - \frac{2}{3}(c-a)\eta + h_3(\varepsilon_V,\eta) \tag{5b}$$

where $h_1$ ($=h_2$) and $h_3$ represent higher-order terms (note that $x_1 = x_2 = x_3$ at $\eta = 0$ as dictated by cubic symmetry). It is worth noting that the linear terms in Eq. (5) suffice for the loss of cubic lattice stability, while the higher-order terms provide a free energy minimum at finite value of the order parameter $\eta$.

The lattice stability of cubic crystal is analyzed through the free energy derivative with respect to strain. For tetragonal lattice deformation, substituting the linear strain dependence in Eq. (5) for $N'$ [ζζ0]-TA2 phonon modes into Eq. (2) yields

$$\frac{\partial F}{\partial \varepsilon_V} = \frac{1}{3}(C_{11}+2C_{12})\varepsilon_V + \frac{1}{3}\chi\frac{k_BT}{\Omega}\left[2f'(x_1)+f'(x_3)\right](c+2a) \tag{6a}$$

$$\frac{\partial F}{\partial \eta} = \frac{2}{3}(C_{11}-C_{12})\eta + \frac{2}{3}\chi\frac{k_BT}{\Omega}\left[f'(x_1)-f'(x_3)\right](c-a) \tag{6b}$$

where $\Omega = V/N$ is primitive cell volume, and $\chi = N'/3N$. There is a marked difference between the two partial derivatives in Eq. (6): for cubic lattice, $\eta = 0$ and $x_1 = x_3$, we obtain $\partial F/\partial \varepsilon_V \neq 0$ while $\partial F/\partial \eta \equiv 0$. The equilibrium volumetric strain $\varepsilon_V^0$ is obtained by solving $\partial F/\partial \varepsilon_V = 0$ at $\eta = 0$:

$$\varepsilon_V^0 = -3\chi\frac{k_BT}{\Omega}f'\left(\frac{\hbar\omega_0'}{k_BT}\right)\frac{c+2a}{C_{11}+2C_{12}} \tag{7}$$

where $\hbar\omega_0'/k_BT = x_0' = x_0 + (c+2a)\varepsilon_V^0/3$. The equilibrium volumetric strain $\varepsilon_V^0$ depends on temperature, i.e., $\varepsilon_V^0 = \varepsilon_V^0[\omega_0(T),T]$, which describes the symmetry-preserving thermal expansion phenomenon. Approximating $\omega_0'$ by $\omega_0$ in Eq. (7), the obtained $\varepsilon_V^0$ is equivalent to Grüneisen model of thermal expansion [12]. Since thermal expansion is an anharmonic effect, Eq. (1) represents an anharmonic theory that takes into account anharmonicity through strain-dependent phonon frequencies $\omega_{\mathbf{k},p}(\varepsilon_{ij})$.

Because $\partial F/\partial \eta \equiv 0$ at $\eta = 0$, in order to analyze the stability of cubic lattice against tetragonal distortion, we consider the second derivative of $F$ with respect to $\eta$ at $\eta = 0$:

$$\frac{\partial^2 F}{\partial \eta^2} = \frac{2}{3}(C_{11}-C_{12}) + \frac{2}{3}\chi\frac{k_BT}{\Omega}f''\left(\frac{\hbar\omega_0'}{k_BT}\right)(c-a)^2 \tag{8}$$

where $\omega_0'$ incorporates the effect of thermal expansion. The cubic lattice loses stability against tetragonal strain when $\partial^2 F/\partial \eta^2 < 0$, which does not require soft mode. Figure 2(a) plots the function $f(x)$ defined in Eq. (3) and its first and second derivatives $f'(x)$ and $f''(x)$. With decreasing $x = \hbar\omega/k_\text{B}T$ (i.e., increasingly softened phonons with decreasing temperature), $f''(x)$ rapidly approaches $-\infty$, reaching the instability condition $\partial^2 F/\partial \eta^2 < 0$ without complete phonon softening. The critical condition $\partial^2 F/\partial \eta^2 = 0$ gives an equation to determine the Curie-Weiss temperature $T_0$ of this cubic→tetragonal phase transition (a first-order transition as shown later):

$$C_{11} - C_{12} = \chi \frac{k_\text{B}T_0}{\Omega} \left| f''\left(\frac{\hbar\omega_0'}{k_\text{B}T_0}\right) \right| (c-a)^2 \tag{9}$$

At temperature $T > T_0$, the phonon energy $\hbar\omega_0'(T)$ is higher, $|f''|$ is smaller, making $\partial^2 F/\partial \eta^2 > 0$, thus the cubic lattice is stable (or metastable when $T_0 < T < T_\text{C}$, $T_\text{C}$ being Curie temperature). With decreasing temperature, phonon energy significantly decreases due to partial phonon softening, and cubic→tetragonal transition occurs at $T_\text{C}$ which is above the martensite start temperature (thus this is a pre-martensitic transformation). It is worth noting that both cubic austenite phase and pre-martensitic tetragonal phase are characterized by the same incompletely softened phonon modes and, more importantly, the phonon dispersion exhibits a continuous dependence on temperature across pre-martensitic transformation; in contrast, martensitic transformation is characterized by discontinuous change in phonon dispersion. It is also worth noting that the elastic constant $C' = (C_{11} - C_{12})/2$ usually undergoes partial softening in cubic austenite phases prior to martensitic transformations [3], which fulfills Eq. (9) at higher $T_0$, i.e., promoting pre-martensitic transformation.

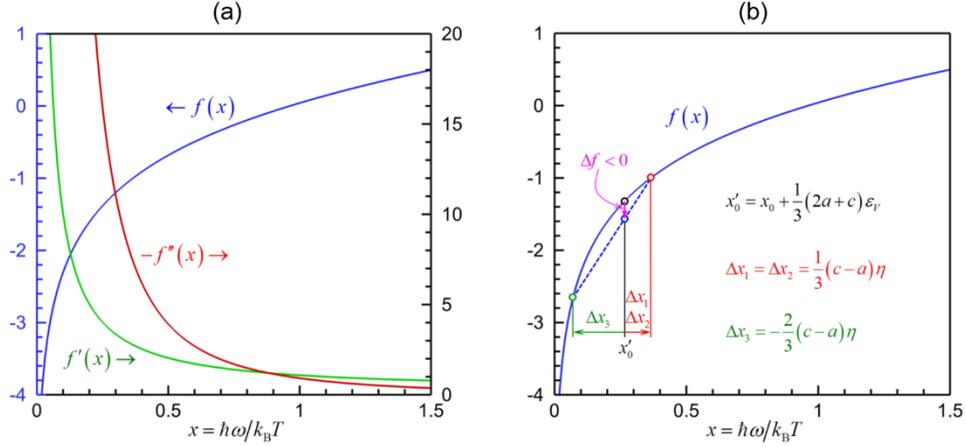

FIG. 2 (color online). (a) Dependences of $f(x)$, $f'(x)$ and $f''(x)$ on $x = \hbar\omega/k_B T$. (b) Schematic illustration of phonon free energy decrease $\Delta f < 0$ upon phonon energy change $\Delta x_i$ associated with tetragonal lattice distortion $\eta$.

The thermodynamic driving force to spontaneous symmetry breaking of cubic lattice can be illustrated schematically, as shown in Fig. 2(b). Keeping only linear strain dependence in Eq. (5) and incorporating thermal expansion into equilibrium cubic lattice (i.e., choosing the state of $\varepsilon_V^0$ and $x_0'$ as reference), a tetragonal lattice strain $\eta$ will induce phonon energy changes $\Delta x_1 = \Delta x_2 = -\Delta x_3/2$ for the 3 respective groups of [ζζ0]-TA$_2$ phonon modes. Because of the negativity of $f''(x)$, the total phonon free energy summed over all 3 groups of the phonon modes will decrease, i.e., $\Delta F_{ph} = \frac{1}{3}\chi \frac{k_B T}{\Omega} f''\left(\frac{\hbar\omega_0'}{k_B T}\right)(c-a)^2 \eta^2 < 0$, as illustrated in Fig. 2(b) where $\Delta F_{ph} = 3\chi k_B T \Omega^{-1} \Delta f$ (this behavior of free energy decrease is analogous to spinodal decomposition, which is also characterized by $f''(c) < 0$ where the free energy $f$ is a function of composition $c$). The decrease in phonon free energy competes with the increase in elastic energy, i.e., $\Delta F_{el} = \frac{1}{3}(C_{11} - C_{12})\eta^2$. When $T < T_0$, the phonon free energy decrease outweighs the elastic energy increase, i.e., $\Delta F = \Delta F_{el} + \Delta F_{ph} < 0$, which is equivalent to $\partial^2 F/\partial \eta^2 < 0$. At $T = T_0$, $\Delta F = 0$, which is equivalent to Eq. (9). It is worth noting that, while the cubic lattice instability is dictated by the linear terms in Eq. (5), to find the equilibrium tetragonal lattice strain $\eta_0$, higher-order terms are required.

The decrease of free energy associated with spontaneous symmetry breaking via development of tetragonal lattice strain $\eta$ is also accompanied by spontaneous redistribution of phonon population among the three groups. The thermal equilibrium occupation number $\langle n \rangle$ of phonons with energy $\hbar\omega$ (or $x = \hbar\omega/k_B T$) is given by Planck distribution function, $\langle n \rangle = (e^x - 1)^{-1}$. An energy change $\Delta x$ leads to a change in phonon occupation number, $\langle \Delta n \rangle = -e^x (e^x - 1)^{-2} \Delta x$. According to Eq. (5), $\Delta x_1 = \Delta x_2 = -\Delta x_3/2$, thus $\langle \Delta n_1 \rangle = \langle \Delta n_2 \rangle = -\langle \Delta n_3 \rangle/2$ within a first approximation, implying $\langle \Delta n \rangle = \langle \Delta n_1 \rangle + \langle \Delta n_2 \rangle + \langle \Delta n_3 \rangle = 0$, i.e., conservation of the total phonon occupation number among the three groups. Nevertheless, the occupation number of phonons is redistributed among the three groups. As a result, the tetragonal lattice is characterized not only by the long-range order parameter (tetragonal strain) $\eta$, but also by the dominant group of phonons with energy $\hbar\omega_3$, which has an excess phonon occupation number

$$\langle \Delta n_3 \rangle - \langle \Delta n_1 \rangle = \exp\left(\frac{\hbar\omega_0'}{k_B T}\right) \left[\exp\left(\frac{\hbar\omega_0'}{k_B T}\right) - 1\right]^{-2} (c-a)\eta$$

over the other two groups of phonons with energy $\hbar\omega_1 = \hbar\omega_2$. Therefore, the spontaneous symmetry breaking of cubic lattice produces three orientational variants of structural domains characterized by the normal vector **n** (i.e., tetragonal $c$-axis), which are both elastic domains of tetragonal lattice and phonon domains of dominant modes. Such elastic-phonon domains can be studied by using three-dimensional single-crystal diffraction experiment to simultaneously measure both Bragg reflection and phonon diffuse scattering that provide complementary information of crystal lattice and lattice vibrations [16]. Thus, in a behavior analogous to spinodal decomposition, the phonon population has an intrinsic tendency to "decompose" into elastic-phonon domains, which is resisted by elastic energy. It is the competition between "spinodal decomposition of phonon population" and elastic energy of crystal lattice that determines the onset of martensitic precursor and pre-martensitic transformation.

With higher-order terms $h_1(\varepsilon_V, \eta)$ and $h_3(\varepsilon_V, \eta)$ present in Eq. (5), the equilibrium values of thermal expansion strain $\varepsilon_V$ and tetragonal lattice strain $\eta$ can be obtained by numerically solving the set of nonlinear equations $\partial F/\partial \varepsilon_V = 0$ and $\partial F/\partial \eta = 0$. As an example to demonstrate the general feature, we incorporate quadratic terms in Eq. (4) and plot the free energy landscape $F(\varepsilon_V, \eta)$ in Fig. 3(a) [17]. The symmetry-preserving thermal expansion corresponds to the local free energy minimum $F_0 = F\left(\varepsilon_V^0, \eta = 0\right)$ defined by $\partial F/\partial \varepsilon_V \big|_{\eta=0} = 0$. The equilibrium tetragonal lattice corresponds to the global free energy minimum $F_{00} = F\left(\varepsilon_V^{00}, \eta_0\right)$. In principle, the free energy function $F(\varepsilon_V, \eta)$ could be approximated by Landau polynomial: expand $F$ into truncated Taylor series in terms of $\varepsilon_V$ and $\eta$, solve $\partial F/\partial \varepsilon_V = 0$ for $\varepsilon_V = \varepsilon_V(\eta)$, substitute $\varepsilon_V(\eta)$ to eliminate $\varepsilon_V$, and obtain $F$ as a polynomial function of $\eta$ only. However, for small value $x$ (e.g., $x = 0.039$ for $\hbar\omega \sim 1\text{meV}$ at $T = 300\text{K}$ typical of incompletely-softened low-energy phonons), the higher-order derivatives $f^{(n)}(x)$ rapidly diverge with increasing order $n$, thus the Landau polynomial does not converge, preventing an analytical approach. Nevertheless, it can be readily shown that the third-order term $\eta^3$ is present in the Landau polynomial, thus the pre-martensitic transformation is a first-order phase transition [1]. The spontaneous symmetry breaking of cubic lattice results in formation of multi-domain structure of tetragonal lattice, producing cross-hatched striation tweed patterns in transmission electron microscopy due to strain contrast, as simulated by phase field microelasticity modeling [18] and shown in Fig. 3(b) [17].

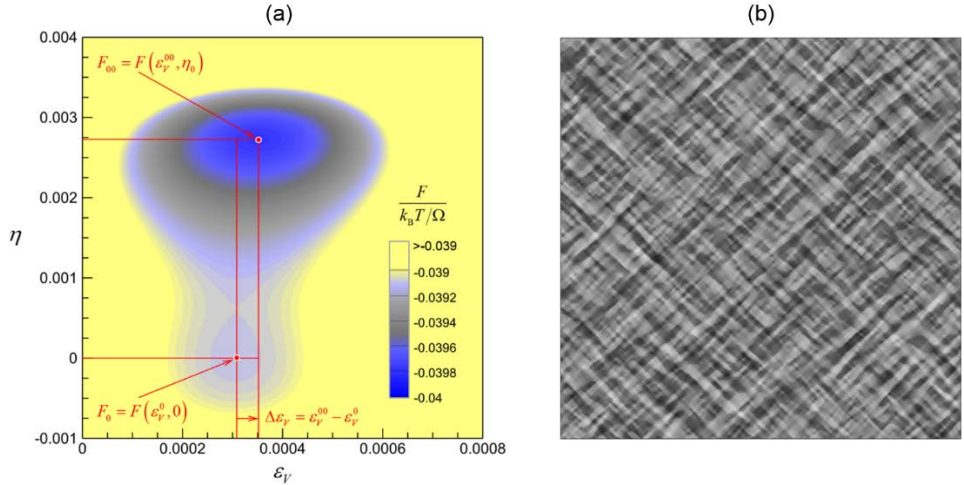

FIG. 3 (color online). (a) Free energy landscape $F(\varepsilon_V, \eta)$ at temperature $T_0 < T < T_C$ to demonstrate first-order pre-martensitic transformation. (b) Tweed structure of lattice coherency strain in multiple elastic-phonon domains of tetragonal lattice.

Based on the phonon-induced behavior of spontaneous symmetry breaking (pre-martensitic transformation), it is natural to explain the ubiquitous precursor anomalies observed prior to martensitic transformations, which have been difficult to explain from conventional phase transition theories [9,10]. The elastic-phonon domains manifest in transmission electron microscopy as tweed patterns, as simulated in Fig. 3(b). The softened phonons produce diffuse scattering in diffraction, where the diffuse "superlattice" spots correspond to the wavevectors of the low-energy phonons [16], e.g., $\zeta \approx 0.33$ at the dip in phonon dispersion curve shown in Fig. 1(a). The existence of elastic domains and domain walls leads to anomalous attenuation of elastic wave propagation observed in pre-martensitic cubic crystals [19]. The phonon occupation numbers in the domains change in response to external stress via internal relaxation process, leading to elastic softening (Le Chatelier's principle [12]), which can be directly observed by Bragg reflection and phonon diffuse scattering [16]. The relaxation time of phonon redistribution leads to frequency dependence of acoustic waves, in particular, much higher values of elastic constants are measured by high-frequency ultrasound than by low-frequency (with longer period relative to phonon relaxation time) and quasi-static measurements [20]. With reference to the symmetry-preserving thermal expansion state (i.e., $\eta \equiv 0$), the symmetry breaking ($\eta \to \eta_0$) brings about additional volumetric strain $\Delta \varepsilon_V = \varepsilon_V^{00} - \varepsilon_V^0$ and free energy change $\Delta F = F_{00} - F_0$, as shown in Fig. 3(a), where the former results in thermal expansion anomaly $\Delta \alpha_V = \partial \Delta \varepsilon_V / \partial T$ and the latter leads to specific heat anomaly (latent heat of pre-martensitic transition) $\Delta C = -T \, \partial^2 \Delta F / \partial T^2$, as observed in martensitic precursors [9,11,21]. Therefore, according to this phonon theory of martensitic transformation precursors, these ubiquitous pre-transitional "anomalies" are natural phenomena like thermal expansion, which are intrinsic behaviors of anharmonic lattice dynamics in cubic crystals that undergo incomplete phonon softening.

Finally, this phonon theory of martensitic precursors can be extended to describe martensitic transformations. The current theory only considers the incompletely-softened phonon modes that are characteristic of the cubic lattice and pre-martensitic tetragonal lattice. In order to describe martensitic transformation, it needs to incorporate the phonon modes that will eventually condense into static lattice modulations (frozen phonons) of the martensite phases. Such further development could lead to a unified theory capable of describing the sequence of pre-martensitic, martensitic and inter-martensitic phase transformations.

This work was supported by US DOE Office of Basic Energy Sciences, Materials Sciences and Engineering Division, Physical Behavior of Materials Program under Award No. DE-FG02-09ER46674.

† Corresponding author: ymjin@mtu.edu

‡ Corresponding author: wangyu@mtu.edu

[1] A. G. Khachaturyan, *Theory of Structural Transformations in Solids* (John Wiley & Sons, New York, 1983).

[2] E. K. H. Salje, *Phase Transitions in Ferroelastic and Co-elastic Crystals* (Cambridge University Press, Cambridge, 1990).

[3] N. Nakanishi, Prog. Mater. Sci. **24**, 143 (1980).

[4] C. Zener, *Elasticity and Anelasticity of Metals* (University of Chicago Press, Chicago, 1948).

[5] P. Souvatzis, O. Eriksson, M. I. Katsnelson, and S. P. Rudin, Phys. Rev. Lett. **100**, 095901 (2008).

[6] W. Cochran, Phys. Rev. Lett. **3**, 412 (1959).

[7] J. A. Krumhansl and R. J. Gooding, Phys. Rev. B **39**, 3047 (1989).

[8] P. A. Lindgård and O. G. Mouritsen, Phys. Rev. Lett. **57**, 2458 (1986).

[9] J. A. Krumhansl, Mater. Sci. Forum **327-328**, 1 (2000).

[10] K. Otsuka and X. Ren, Prog. Mater. Sci. **50**, 511 (2005).

[11] A. Planes, E. Obradó, A. Gonzàlez-Comas, and L. Mañosa, Phys. Rev. Lett. **79**, 3926 (1997).

[12] L. D. Landau and E. M. Lifshitz, *Statistical Physics* (Pergamon Press, Oxford, 1980).

[13] A. Zheludev, S. M. Shapiro, P. Wochner, A. Schwartz, M. Wall, and L. E. Tanner, Phys. Rev. B **51**, 11310 (1995).

[14] A. Zheludev and S. M. Shapiro, Solid State Communications **98**, 35 (1996).

[15] The material parameters characterizing $\omega_{\mathbf{k},p}(\varepsilon_{ij})$, e.g., the Taylor expansion coefficients, can be determined by first principles density functional theory computations at finite temperature using self-consistent *ab initio* lattice dynamics approach [5].

[16] Y. M. Jin, Y. U. Wang, Y. Ren, and F. Ma, "Three-Dimensional Phonon Diffuse Scattering and Bragg Reflection Study of Domain Phenomena in Martensitic Precursors" (to be published).

[17] See Supplemental Materials for details of material parameters and phase field microelasticity modeling used in our computation and simulation in Fig. 3.

[18] Y. U. Wang, Y. M. Jin, and A. G. Khachaturyan, Acta Mater. **52**, 1039 (2004).

[19] T. R. Finlayson, Metall. Trans. A **19**, 185 (1988).

[20] M. Wuttig (unpublished data, private communication).

[21] M. Liu, T. R. Finlayson, T. F. Smith, and L. E. Tanner, Mater. Sci. Eng. A **157**, 225 (1992).